**Towards physical principles of biological evolution**

Mikhail I. Katsnelson[1,*], Yuri I. Wolf[2], Eugene V. Koonin[2]

[1]Radboud University, Institute for Molecules and Materials, Nijmegen, 6525AJ, Netherlands

[2]National Center for Biotechnology Information, National Library of Medicine, Bethesda, MD 20894

*For correspondence: M.Katsnelson@science.ru.nl

**Abstract**

Biological systems reach organizational complexity that far exceeds the complexity of any known inanimate objects. Biological entities undoubtedly obey the laws of quantum physics and statistical mechanics. However, is modern physics sufficient to adequately describe, model and explain the evolution of biological complexity? Detailed parallels have been drawn between statistical thermodynamics and the population-genetic theory of biological evolution. Based on these parallels, we outline new perspectives on biological innovation and major transitions in evolution, and introduce a biological equivalent of thermodynamic potential that reflects the innovation propensity of an evolving population. Deep analogies have been suggested to also exist between the properties of biological entities and processes, and those of frustrated states in physics, such as glasses. Such systems are characterized by frustration whereby local state with minimal free energy conflict with the global minimum, resulting in "emergent phenomena". We extend such analogies by examining frustration-type phenomena, such as conflicts between different levels of selection, in biological evolution. These frustration effects appear to drive the evolution of biological complexity. We further address evolution in multidimensional fitness landscapes from the point of view of percolation theory and suggest that percolation at level above the critical threshold dictates the tree-like evolution of complex organisms. Taken together, these multiple connections between fundamental processes in physics and biology imply that construction of a meaningful physical theory of biological evolution might not be a futile effort. However, it is unrealistic to expect that such a theory can be created in one scoop; if it ever comes to being, this can only happen through integration of multiple physical models of evolutionary processes. Furthermore, the existing framework of theoretical physics is unlikely to suffice for adequate modeling of the biological level of complexity, and new developments within physics itself are likely to be required.

**Introduction**

How are living organisms different from inanimate matter? There are obvious answers in terms of the chemical composition and structure (at least as far as the only known case in point, namely, life on earth, is concerned), but when it comes to the central processes in the evolution of life, the distinction is far less obvious. In the tradition of Darwin-Wallace, it is tempting to posit that life is defined by evolution through the survival of the fittest [1-5]. However, the uniqueness of this process to life could be questioned because the entire history of the universe consists of changes where the most stable (the fittest) structures survive. The process of replication itself is not truly unique to biology either: crystals do replicate. On the macroscopic scales of space and time, however, life clearly is a distinct phenomenon. To objectively define the features that distinguish life from other phenomena that occur in the universe, it seems important to examine the key processes of biological evolution within the framework of theoretical physics [6, 7].

Arguably, the central feature that distinguishes modern physics from other areas of human endeavor is the distinct relationship between theory and experiment whereby research programs are shaped by testable theoretical predictions. In general, modern biology is not a theory-based science in the sense physics is. There is, however, a major exception, namely, population genetics, a formalized field of biology that is structured effectively as an area of theoretical physics, akin primarily to statistical thermodynamics [8-11]. To wit, population genetic formalisms have been highly efficient in immunology [12, 13] and cancer biology [14-17], perhaps, suggesting that further expansion of theory in biology could be possible and productive. Modern theoretical physics is a tightly interconnected area in which widely different fields are intertwined. At present, population genetics or any other direction of theoretical biology is not part of that network. It can be argued that this disconnect is not the optimal state of affairs because many areas of theoretical physics could inform and stimulate theoretical developments in biology.

Then, one is bound to ask: is modern physics sufficiently rich to encompass biology? This question, posed in various forms (in particular, "is biology reducible to physics?"), has a long and rather torturous history (e.g. [18, 19]). Without going into historical or philosophical details, we dismiss any suggestion that life could follow some special laws of "biological" physics instead of the generally established ones. Quantum mechanics, in particular, is generally valid and applies to living organisms just as well as to any other form of matter. The problem is that this powerful theory that, in a sense, can be considered a "theory of everything" does little if anything to explain biological phenomena [20, 21]. Certainly, quantum mechanical calculations can be useful for analysis of biochemical reactions, but they do nothing to help us understand evolution. Therefore, it has been suggested that the physical concept that could be pivotal for the theoretical description of biological phenomena is emergence, i.e., collective behavior of large ensembles that is qualitatively distinct from the behavior of the constituent entities: "More is different" as aphoristically formulated by Anderson [20-25] .

One of the most fundamental and difficult problems in biology is the origin and evolution of the elaborate order and enormous complexity of living organisms. Complexity is one of the most challenging concepts in all of science that resists all-encompassing definitions [26]. Indeed, the most useful definitions of complexity appear to be context-specific. In biology, complexity is relevant, at least, at the levels of genomes, organisms, and ecosystems [27, 28]. The genomic complexity may be meaningfully defined as the number of nucleotide sites that are subject to selection and thus carry biologically relevant information [29-31], although such a definition misses other important sources of genome-level complexity, such as alternative transcription initiation and alternative splicing in eukaryotes. Organismal and ecological complexity is usually perceived as the number of distinct constituent parts and/or levels of hierarchy in the respective systems [32]. Regardless of the exact definitions, stably maintained, evolving high level of complexity is a distinctive feature of life and a major challenge to theory.

The most traditional interface between physics and biology is biophysics, i.e. study of the properties of the structure and dynamics of biological macromolecules as well as cellular and organismal structures and functionality using physical approaches. Various directions in biophysics have been productive and successful over many decades [33]. There is, however, a distinct, complementary area of interaction between physics and biology whereby physical theory is used to describe, model and analyze biological processes, in particular, evolution at the population level [4, 6, 7]. Parallels between thermodynamics and statistical mechanics, on the one hand, and population genetics, on the other hand, have been already invoked by the famed statistician and a founding father of population genetic theory, Ronald Fisher, as early as the 1920s [34], and have been explored and extended in recent years [8, 10, 11]. In different forms, formalisms from statistical mechanics have been increasingly employed to model biological evolution. Among others, a notable application is the use of percolation theory for analysis of evolution on fitness landscapes [35-37]. The ultimate goal of this injection of physics into evolutionary biology appears highly ambitious: nothing less than development of a physical theory of biological evolution, or even reshaping of biology as an area of physics [6, 7]. Obviously, this type of an overarching research program, even if feasible in principle, cannot be realized in a single, clean sweep. It can only progress one step at a time, by modeling various evolutionary processes using ideas and mathematical apparatus from theoretical physics and hoping that eventually, it becomes possible to combine such models into a coherent theoretical framework.

In this article, we discuss several aspects of biological evolution where theoretical insights, coming primarily from condensed matter physics, appear possible. We submit that physical theory can make non-trivial contributions to the current understanding of evolution but new theoretical developments within physics itself are likely to be required to fully account for the emergence and evolution of the level of complexity that is characteristic of biological systems.

**The correspondence between thermodynamics and population genetics, and major evolutionary transitions**

Although the existence of parallels between statistical mechanics and population genetics has been realized early in the history of the latter, the detailed correspondence has been derived by Sella and Hirsch in 2005 [8] and further developed by Barton and colleagues [10, 11] (Table 1)

Perhaps, the most notable equivalence is that between the effective population size and inverse temperature ($N_e \sim 1/T$). In a full analogy with physical systems, evolution is effectively deterministic at low $T$ such that an infinite population (an abstract construct often used in population genetic research) is equivalent to 0 K, that is, to the ground state of a physical system. Whereas the latter is usually unique (non-degenerate), in an infinite-size population, the selection pressure is so strong that only one, globally optimal configuration survives, at least in the infinite time limit. In contrast, at high $T$ (small populations), evolution becomes a stochastic process that is dominated by fluctuations (or genetic drift, in the language of population genetics). This stochastic regime involves a multiplicity of allowed evolutionary trajectories (in other terms, valleys of low fitness in fitness landscapes can be crossed) and accordingly provides for innovation and emergence of biological complexity.

Here, we take a more general approach and draw parallels between parameters of the evolutionary process and quantities from phenomenological thermodynamics (rather than statistical mechanics). Formally, it appears natural to introduce a quantity $I$ that is analogous to thermodynamic potential and changes during evolution:

$$dI = dt(dS/dt)/N_e \qquad (1)$$

which seems to have a clear biological meaning. Here, $S$ is evolutionary entropy [4, 31] that is calculated as follows:

$$S = \sum_{i=1}^{L} S_i = -\sum_{i=1}^{L} \sum_j f_{ij} \log f_{ij} \qquad (2)$$

where $S$ is the total entropy of the alignment of $n$ sequences of length $L$; $S_i$ is the per site entropy and $f_{ij}$ are the frequencies of each of the 4 nucleotides ($j=A,T,G,C$) or each of the 20 amino acids in site $i$. Equation (2) is equivalent to the classic Shannon formula [38] except that, instead of applying it "horizontally", i.e. to a single sequence, it is applied "vertically", i.e. to an alignment of homologous sequences, hence "evolutionary entropy". In the definition of evolutionary entropy (Eq. 2), genetic changes in small and large populations are taken with the same weight. However, evolutionary innovations occur mostly in small populations where selection is weak and more variants have a chance to survive [39, 40]. The quantity $I$ (Eq. 1) reflects this trend and has the meaning of *evolutionary innovation potential*. Although the required calculations could be involved, the values of $I$ and $dI$ can be extracted from reconstructions of genome evolution and compared to other features, such as genome size, various measures of genome complexity and selection pressure.

This line of reasoning is fully compatible with the concept of evolution of biological complexity that was developed by Lynch purely from population genetic considerations [39-41]. More specifically, such is the origin of complexity of multicellular organisms that is manifest on both the genomic and the organismal levels [40, 42]. Clearly, the genomes of multicellular organisms, such as animals and plants, with their haphazard organization (sparse genes, coding sequences interrupted by introns, lack of tight clustering of functionally related genes), are typical high temperature objects that are characterized by disorder, and hence, complexity. Conversely, genomes of prokaryotes and viruses, with the characteristic high gene density and operonic organization [43], are much more ordered, which corresponds to low temperature.

Discrete levels of selection lead to discrete levels of biological complexity. This modality of biological evolution is encapsulated in the concept of major transitions in evolution (MTE) developed by Szathmary and Maynard Smith [44, 45]. In each MTE, the units of the preceding level form ensembles that become new units of selection. The key major transitions include the origin of cells from pre-cellular life forms (even if the latter remain poorly understood), origin of eukaryotes via endosymbiosis, origin of multicellularity (which occurred independently on several occasions), and origin of animal eusociality and superorganisms in plants and fungi. Within the framework of the correspondence between statistical physics and population genetics (Table 1), the MTE can be readily interpreted as analogs of the first-order phase transitions [46] (Figure 1) . Usually, they are considered for systems in a thermal bath, so that temperature remains constant but there is a jump in entropy related to the latent heat of transformation. In the context of biological evolution, temperature corresponds to the inverse population size (Table 1) and, obviously, changes during MTE. Indeed, the transitions lead to increased size of individuals and, accordingly, energy flux at the new level of organization (and selection): for example, the volume of a eukaryotic cell is about 1000 fold greater than that of a typical prokaryotic cell. Effective population size is well known to scale inversely with the organism size, so the MTE are accompanied by abrupt rise in the evolutionary temperature. An entropy-related quantity that, perhaps counter-intuitively, remains roughly constant during evolution, and through the MTE, is *evolutionary information density* :

$D(N) = 1 - S/N$ (3)

where $S$ is the evolutionary entropy of Eq. (1), and $N$ is the total length (number of sites) of a genome (simplified from [31]).

Thus, the evolutionary transitions appear to be analogous to *adiabatic* first-order transitions, with evolutionary information density and evolutionary temperature (effective population size) being thermodynamically coupled variables. Coming back to physics, ultracold gases provide an excellent example of thermodynamics at a constant entropy [47, 48]. More formally, first-order phase transitions at constant temperature are determined from the equality of chemical potentials $\mu$ (Gibbs potentials per particle) of different phases. The differential of the chemical potential changes with temperature change for a given entropy value per particle is $d\mu = -sdT$ where $s$ is the entropy density. For adiabatic transitions (constant entropy), the corresponding quantity is the energy per particle $e$, with the differential $de = Tds$. Given that, within our analogy, $D(N)$ corresponds to $s$ and $T$ corresponds to $1/N_e$, our "innovation potential" (1) turns out to be the "thermodynamic potential". Thus, first-order phase transitions are characterized by temperature jumps, which is exactly what happens at MTE within the framework of the evolution-to-thermodynamics mapping (Table 1 and Figure 1).

**Life, glasses and patterns: Frustrated systems and biological evolution**

As first clearly introduced in the spin-glass theory by Edwards and Anderson [49], modern physics considers glass to be a distinct state of matter that is intermediate between equilibrium and nonequilibrium [50-53]. A characteristic property of glass is aging, or structural relaxation. Suppose we measure a specific property of an equilibrium phase, liquid or solid, e.g. the resistivity of metal (or liquid metal). "Equilibrium" means that, when the measurement is repeated after a thermal cycle (slow heating and cooling down to the initial temperature), we obtain the same value of the resistivity. In glass, the measured value would slowly change from measurement to measurement. The potential energy relief (or landscape, to use a term with biological connotations) for glass is a function with many (asymptotically, infinitely many) local minima separated by barriers with an extremely broad energy distribution. Each local minimum represents a metastable state. During its thermal evolution, the system slowly moves from one minimum to another. Importantly, the glass state is non-ergodic [50-53]. The state of the glass is characterized by an "order parameter" with continuously many components, labeled by a real number $x \in (0,1)$ [54]. This number can be represented as an infinite, non-periodic binary fraction, such as 0.10001110…, where 0(1) corresponds to the choice of bifurcation on the complex energy relief when cooling down from the equilibrium liquid state. This feature can be conceived of as a specification of the aperiodic crystal concept introduced by Schrödinger in his famous book [55]. A major distinction is that glasses are not only aperiodic but also non-ergodic, a feature that results in an evolutionary process. The relevance of the concept of glassiness in biology has been emphasized by Laughlin and colleagues [20, 21]. However, the defining features of life, namely replication with selection, seem to go beyond simple glassy behavior: the potential relief of glasses appears too flexible and too generic to model biological evolution. Glass displays effectively infinite variability, whereas life is based on discrete forms, such as genomes with defined sequences and distinct, extended intervals of stability (see the discussion of evolutionary transitions below).

One of the formal criteria of the glass state is "universal flexibility" [56]. Omitting some important but purely technical details, it can be described as follows. Consider a configuration (of spins, atomic positions, dipolar moments and other parameters) that is characterized by a function $\phi(x)$ where $x$ is $d$-dimensional vector characterizing a position in space (in most physical applications, $d = 2$ or 3). The energy of this configuration is given by its Hamiltonian $H[\phi(x)]$ and free energy

$$F = -T \ln \int D\phi \exp(-H[\phi(x)]/T) \tag{5}$$

where $T$ is the absolute temperature (we put Boltzmann constant equal to one) and $\int D\phi$ represents summation over all possible configurations. Let us add interaction with another configuration $\sigma(x)$:

$$H[\phi(x)] \to H_g[\phi(x)] = H[\phi(x)] + \frac{g}{2} \int dx [\phi(x) - \sigma(x)]^2 \tag{6}$$

and calculate the free energy $F_g$ replacing $H[\phi(x)] \to H_g[\phi(x)]$ in Eq.(1). Then, let us consider two transitions: thermodynamic limit $V \to \infty$ where $V$ is the volume of the system and the limit of infinitely weak coupling $g \to +0$. If these limits do not commute, i.e.

$$\lim_{g \to +0} \lim_{V \to \infty} \frac{F_g}{V} \neq \lim_{V \to \infty} \lim_{g \to +0} \frac{F_g}{V} \tag{7}$$

for macroscopically large number of configurations $\sigma(x)$, then, the system is glass. Physically, this means that the energy relief for the glass is reminiscent of a "universal mapping function", so that for many $\sigma(x)$, there exists a part of the relief that is minimized by the choice $\phi(x) = \sigma(x)$.

The original glass concept was developed for disordered systems with some type of randomness in the interatomic interactions. Actually, such randomness is not essential as clearly demonstrated in the concept of *self-induced glassiness* [57-59]. It turns out that some systems satisfy the criterion of Eq. (7) without any randomness but, necessarily, in the presence of frustrations caused by competing interactions on different spatial scales.

For the case when $\sigma(x)$ is simply one specific function, equation (7) is equivalent to the condition of *spontaneously broken symmetry* in Landau theory of second-order phase transitions [46, 60], with $h(x) = g\sigma(x)$ playing the role of external field conjugated to the order parameter $\phi(x)$. Conceivably, for some systems, the criterion (6), (7) can be satisfied neither for an "almost arbitrary" function $\sigma(x)$ as in glasses nor for a single function as in conventional second-order phase transitions, but for a sufficiently rich but limited set of functions. Such systems would spontaneously "glue" to selected configurations from some "library" to form a complex but not completely chaotic pattern. Such patterns might yield better models for biological phenomena than classical glass. In other words, we assume that there is a number of *discrete*, separated "attractors" and that the energy landscape of the system consists of glassy parts separated by gaps. This model immediately invokes an analogy with pattern recognition that has been successfully studied using the spin-glass theory [50]. A clear analogy in evolutionary biology is a fitness landscape with elevated areas of high fitness, where an evolving population can travel either upwards, under the pressure of selection, or horizontally in a (quasi)neutral evolutionary regime, separated by valleys of low fitness that can be crossed only by genetic drift (thermal fluctuation) as discussed above [35]. A less obvious but potentially important point concerns the concept of "order from disorder" [61-63]. Consider a frustrated system with competing interactions. Such a system can assume many states with the same energy that are characterized by different types of ordering. The system cannot choose between these states because they are completely degenerate in terms of total energy and so remains disordered. *Frustration* and *competing interactions* are key concepts in the theory of (spin) glasses [50-53] . Generally, the states with the same ground-state total energy have different excitation spectra over the ground states and therefore different entropies at finite temperature. As a result, free energies of different ordered states become different, the degeneracy is broken, and the system becomes ordered by "choosing" one of the competing ordered states. In this case, entropy creates order from disorder rather than destroying it although the second law is by no means violated. Thus, ordering that can be equated with "meaningful" complexity results directly from frustration in these relatively simple physical systems.

The concept of frustration as the source of complexity seems to be directly relevant for understanding biological evolution. Conflicts between genetic elements and biological entities at different levels of organization that result in frustration permeate all of biology [64, 65](Figure 2 and Table 2). Arguably, the most obvious form of these conflicts is the competition between evolutionary strategies of parasites and hosts [64, 66-69]. Genetic parasites with different reproduction modes, including viruses, plasmids, and transposons, are associated with (nearly) all cellular life forms [4, 69, 70]. The emergence and persistence of such parasites appears to be an intrinsic feature of biological replicator systems because it can be shown that parasite-protected systems are inherently evolutionarily unstable [71]. Frustration caused by intergenomic conflicts drives the evolution of biological complexity [72]. Indeed, computer simulations under a wide range of conditions consistently show that, in a well-mixed replicator system, parasites overwhelm the hosts and eventually cause collapse of the entire system [73-76]. In such simulations, compartmentalization stabilizes the system, being also a path to diversification and evolution of complexity. Once again, the outcome of such modeling studies represents patterning, a typical consequence of frustration in glass-like states (Figure 2). Apart from compartmentalization, host-parasite conflicts drive the evolution of versatile defense systems in the host and counter-defense systems in parasites, which is another prominent manifestation of biological complexity [66, 77-79].

Notably, the conflicts between hosts and parasites are resolved in many different ways, into stable evolutionary regimes that span the entire range between highly aggressive parasites, such as lytic viruses, that kill the host and move to the next one, and cooperative elements, such as plasmids, that provide beneficial functions to the host [68, 69]. This diversification of host-parasite interactions is an important part of biological complexity at the level of ecosystems and the entire biosphere.

The frustration caused by host-parasite conflicts is an important driver of the MTE [72](Table 2). A nearly ubiquitous anti-parasite strategy in virtually all cellular life forms is programmed cell death (PCD), i.e. altruistic suicide of infected cells that prevents virus spread in a population [79-84]. PCD becomes an efficient defense strategy only in cellular aggregates and thus is likely to have been one of the key factors in the evolution of multicellularity [85, 86].

The origin of eukaryotes itself clearly was, to a large extent, driven by frustration, in this case, the conflict between the protomitochondrial endosymbiont and its host, most likely, an archaeon related to Lokiarchaeota [87-93]. This conflict seems to have played out at several levels including an onslaught of selfish genetic elements from the evolving endosymbiont on the host genome that most likely drove the evolution of the exon-intron architecture of eukaryotic genes, one of the central features of the genomic complexity in eukaryotes [89, 94]. The multiple, entangled competing interactions between the parasite turning into an endosymbiont and the host led to numerous innovations in the cellular organization of the emerging eukaryotes, conceivably, through the extreme population bottleneck during eukaryogenesis, which was caused by the host-parasite frustration [95-97]. The frustration was resolved by the formation of the stable symbiotic association, the eukaryotic cell, but the conflict lingers, e.g. in the form of mitochondrial diseases [98] and frequent lysis of mitochondria that in some organisms results in insertion of mitochondrial DNA into the host genome [99].

Host-parasite coevolution that involves both arms race and cooperation is a fundamentally important but far from the only manifestation of frustration in biological systems. Various forms of frustration are detectable at all levels of biological organization, from macromolecules to the biosphere, and at the heart of each major transition (Table 2). The case of multicellularity is particularly transparent. Frustration caused by the conflict between the selection pressures at the cellular and organismal levels is an intrinsic feature of the evolution of multicellular life forms [100, 101]. The stable resolution of this frustration involves control over cell division, providing for the evolutionary stability of multicellular organisms, but an alternative, also common solution is cancer [102-104].

Sexual reproduction involves a different level of conflict [105, 106], and eusociality, obviously, entails another [107]. It does not seem to be much of a stretch to posit that frustrated states underlie the entire course of the evolution of life.

**Percolation and criticality as the basis and condition of tree-like evolution**

Evolution can be described as percolation [108, 109] on a multidimensional fitness landscape (or a dynamic "seascape") [35-37, 110, 111]. The dimensions can correspond to selectable traits or genes and thus can be in the thousands. The simplest image is a landscape with mountain ridges and plateaus that are partially covered with water. On such a landscape, paths that go under water (below the survival threshold) never continue, whereas the accessibility of the paths above the water level depends on the evolutionary temperature (or effective population size) as outlined above. Applied to biological evolution, this description appears quite complicated [37]. However, the problem becomes much simpler if we consider a *critical* percolation cluster that consists of paths that follow the shores, that is, the lines of *minimal* fitness that is necessary for survival (Figure 3). There are good reasons to believe that actual evolution does not deviate far from such paths due to the cost of selection [112-116]. In this case, it is, paradoxically, the multidimensionality of the parameter space that drastically simplifies the problem. Percolation clusters in two- or three-dimensional spaces have complicated structures, with many doubling channels and dead ends [108, 109], and non-trivial fractal properties [117-119]. In contrast, it has been mathematically proven [120-122] that, for a space of sufficiently high dimensionality $d$ (in the simplest cases, $d>5$), the structure of the critical percolation cluster is approximated by a tree (or Bethe lattice). In other words, the cluster has a simple tree structure, without double paths or dead ends and is an "optimal" simple line. An important consequence of the tree-like structure of the critical percolation cluster in a high dimensionality space is that, typically, there is only one passable route between any two points on such landscapes.

It seems plausible that the multidimensionality of the biologically relevant fitness landscapes is the cause of the tree-like trend that is readily decipherable in the evolution of all cellular life forms [123, 124]. At the level of phylogenomics, this trend reflects the coherence between the topologies of the phylogenetic trees for different individual genes [125], notwithstanding the extensive horizontal gene transfer in prokaryotes that has been invoked to question the very validity of the tree-like character of evolution [126-130]. This is not true for small viruses with only a few genes, in which case evolution can be adequately represented only by a network [131-133]. Thus, although the exact theory remains to be developed, the percolation perspective strongly suggests that coherent evolution of a gene core resulting in the tree-like structure of organismal evolutionary trajectories – and the possibility of speciation – is a consequence of the high dimensionality of the fitness landscapes. As such, tree-like evolution appears to be a fundamental property of life that stems from basic physical principles.

**The genotype-phenotype mapping and selection as measurement**

A necessary condition of evolution is the separation of the genotype and the phenotype, in the simplest case, a replicator and a replicase, and the existence a genotype-to-phenotype mapping [4, 134]. The evolving genotype (genome) can be represented as a discrete state vector, whereas the phenotype evolves in a continuous space [135]. Thus, the genotype and the phenotype can be represented as two pattern-like phases (see above) with vastly different numbers of degrees of freedom. The mapping of the genotype onto the phenotype, or in other words, the distribution of the phenotypic effects of genotype mutations is a non-trivial problem. A self-evident condition of evolution is the actual existence of mapping, i.e. a feedback from the phenotype to the genotype, such that at least some mutations result in heritable phenotypic changes affecting the fitness of the organism. This type of mapping between two distinct glass-type phases with different properties presents a problem that, to our knowledge, so far has not been addressed in physics and appears to require a special language. Although the current framework of statistical physics seems not to be immediately suitable for analysis of objects that simultaneously exist on two coupled levels, such situations are not completely new in modern physics. Here we discuss some conceptual similarities with the procedure of measurement in quantum physics, which deal with two types of processes, namely, smooth evolution interrupted by some "projections". The phenotype can be regarded as a gauge that measures the fitness of the genotype, and in that regard, the interaction of the phenotype with the genotype is closely analogous to measurement (Von Neumann prescription) in quantum mechanics [136](Figure 4).

The concept of *measurement* plays the central role in quantum mechanics [136-140]. The physical framework of this concept was developed mainly by Bohr in his complementarity principle [141] and formalized by von Neumann in his mathematical theory of quantum measurements [136]. According to this approach, existence of *classical* objects, namely measurement devices, is postulated. The state of a quantum system is characterized by the wave function (state vector) $|\Psi\rangle$ or, equivalently, by the density matrix which, for an isolated quantum system, is simply a projector operator into this state: $\hat{\rho} = |\Psi\rangle\langle\Psi|$. For an isolated quantum system, the system dynamics is described by *unitary evolution* $\hat{\rho}(t) = \exp(it\hat{H}/\hbar)\hat{\rho}(0)\exp(-it\hat{H}/\hbar)$ where $\hat{H}$ is the Hamiltonian (which, for simplicity, is supposed to be time-independent), $t$ is time and $\hbar$ is the Planck constant. During this evolution, entropy $S(t) = -Tr\hat{\rho}(t)\ln\hat{\rho}(t)$ remains equal to zero, thus, this process is in principle reversible. After the interaction with the measurement device, the density matrix abruptly becomes diagonal in some basis $\{|n\rangle\}$ depending on the specific nature of the device:

$$\hat{\rho} \to \sum_n (\langle n|\hat{\rho}|n\rangle)|n\rangle\langle n| \tag{8}$$

This transition is known as *von Neumann prescription* and is accompanied by entropy increase. During the measurement, the information contained in off-diagonal elements of the density matrix in the basis $\{|n\rangle\}$ is irreversibly lost.

This duality between the purely quantum unitary evolution and projection by measurement that involves classical devices is considered by many physicists as unsatisfactory, and there have been various efforts to justify the appearance of classical objects in the quantum world and to derive von Neumann prescription from the Schrödinger equation (see, e.g., Refs. [142-144]). However, there is no accepted solution to the problem. One approach involves information theory whereby measurement is perceived as a fundamental concept underlying quantum mechanics (see, e.g., Refs. [145, 146]) where, in contrast to the conventional view, the Schrödinger equation is derived from analysis of the measurement procedure.

There seems to exist a striking analogy between measurement in quantum physics and the process of biological evolution [147, 148]. One can consider evolution of the genotype under the laws of classical (or quantum) physics being interrupted by the feedback from the phenotype in the form of selection. Had selection on mutations been the only factor of evolution, the process would reduce to adsorption of phase trajectories at the borders of some regions of the genotype space (those that yield phenotypes incompatible with survival). Then, under the action of Müller's ratchet, all states would be, sooner or later, eliminated [149-151]. However, genetic drift and especially horizontal gene transfer, including sex, open tunnels into regions of the genotype space that are inaccessible to the smooth evolution [152-154]. Thus, future "biology as theoretical physics" should probably deal with two types of dynamics similar, respectively, to unitary evolution and measurement in quantum mechanics, with the selection of phenotypes playing the role of "measurement". Before the measurement is performed, the quantum system evolves along all possible paths, as explicitly embodied in Feynman's path integral formulation of quantum mechanics [155]. Measurement induces the wave function collapse such that, of all evolutionary trajectories of the system, only those are chosen that pass through a given point. Similarly, any genome sequence is formally possible and indistinguishable from others until its fitness is measured by selection at which point genotypes corresponding to low fitness are eliminated. Evolution works as Maxwell Demon [156, 157] selecting genotypes by some external (with respect to the genome sequence) criteria. Crucially, this cannot be done for free. According to the Landauer principle (ultimately stemming from the Second Law of Thermodynamics), obtaining information on any system incurs a cost of at least $kT\ln2$ J/bit, where $k$ is Boltzmann constant and $T$ is temperature [157, 158]. This cost of measurement appears to be equivalent to the cost of selection that was first introduced by Haldane [122], which puts limits on adaptive evolution analogous to the limits on measurement precision.

We would like to emphasize that the analogy between theory of evolution and quantum physics discussed here is focused solely on the role of measurement which seems to be crucially important in both cases [147]. This probably will affect the mathematical form of the prospective physical theory of evolution which can be expected to include formalism similar to the von Neumann projection. A necessary disclaimer is that this approach has nothing to do with "quantum biology", i.e. attempts on direct application of quantum physics to biological processes that is hardly justified.

**Concluding remarks**

A "general physical theory of biology" might be a vacuous dream but it seems possible to naturally describe key evolutionary processes in the language of statistical physics. Here we outline four areas in evolutionary biology where analogies with models from theoretical physics appear natural and might be constructive: i) correspondence between quantities that characterize genome evolution and those that are used to describe the evolution of simple physical systems in phenomenological thermodynamics, ii) competing interactions and frustrated states, analogous to those in the theory of striped glasses, as one of the key factors of biological evolution, iii) evolution as percolation and criticality of the percolation cluster as the condition of tree-like evolution, iv) genotype-phenotype mapping as correspondence between two glass-like phases and selection as a form of measurement. These themes could appear disjoint, and indeed, they hardly form a coherent theoretical framework. Nevertheless, they all share a unifying, dominant thread, namely, the emergence of new levels of complexity from simple physical principles. We find it conceivable that the development of the future physical theory of evolution proceeds along these lines.

It is now commonly accepted that random processes play essential roles in evolution and that biological complexity is, at least partially, driven by fluctuations. The use of statistical physics is therefore natural. However, we should not go too far. Natural selection and adaptation are essential to biological evolution as well, and to incorporate these phenomena into a framework of physical theory, the existing apparatus of statistical physics probably requires amendments. Here, we tried to speculate on what kind of modifications might be needed. Emergent phenomena that are inherent to the theory of patterns, glasses and other condensed matter states are also central in biology. However, special principles, not yet developed in statistical physics, appear to be required for a physical theory of the genotype-phenotype separation and mapping that comprise the cornerstone of evolution. Biological evolution by no means defies any laws of physics but the emergent biological phenomena appear to call for extension of physics itself. Biological entities and their evolution do not simply follow the "more is different" principle but, in some respects, appear to be qualitatively different from non-biological phenomena, indicative of distinct forms of emergence that require new physical theory. The difference between biology and physics (at least as we know it) is not that "nothing in biology makes sense except in the light of evolution" [3] whereas in physics "everything does". The latter statement does not actually appear to be true outside of the quantum physics confines because the entire universe certainly can be properly understood only in the light of its evolution over 13.8 billion years. Following the analogy outlined above, in biology as in physics, measurement generates the arrow of time and necessitates evolution. However, biological evolution has substantial special features, some of which we tried to capture here, in particular, by applying concepts of condensed matter physics, such as frustration and percolation, to central processes of biological evolution. Evidently, the analysis and discussion presented here are only prolegomena to the sustained, concerted effort, which is required to unite biology and physics.

Figure legends

Figure 1. Major evolutionary transitions as adiabatic first-order transitions with constant entropy and changing temperature.

The 3 panels schematically show the changes in effective population size (top panel), evolutionary information density (middle panel) and the evolutionary innovation potential introduced here (bottom panel) at a MTE that is shown by the vertical dotted line.

Figure 2. Frustration as a key driver of biological evolution.

The left side illustrates a typical frustrated state in condensed matter physics which is exemplified by spin interaction in a glass-like system. The right side illustrate conflicts and frustration in biological evolution exemplified by host-parasite interaction.

Figure 3. Critical percolation and tree-like evolution.

A, Ancestor; D1-5, descendants. The dotted line shows the evolutionary trajectory from the ancestors to descendants. In critical percolation, these trajectories include no loops and are tree-like.

Figure 4. Evolution as measurement.

The genotype is shown by a 0/1 string for generality.

Table 1

**The correspondence between the key variables of statistical physics and evolutionary biology**

|  | **Biological analog** | |
| --- | --- | --- |
| **Thermodynamic variable** | Sella and Hirsh [8] | This work |
| Inverse temperature, $\beta=1/T$ | Effective population size $N_e$ | |
| Entropy per particle | Derived from the free fitness expression | Evolutionary information density: $D(N)= 1-H/N$ (see text) |
| Free energy Hamiltonian | Minus log of fitness | - |
| Thermodynamic potential | Derived from the Hamiltonian by Gibbs formula | Evolutionary innovation potential: $dI= dt(dH/dt)/N_e$ |

Table 2

**Frustrated relationships in biological evolution**

| System | Frustration-producing elements (competing interactions) | Evolutionary consequences |
|---|---|---|
| Proteins | Hydrogen and Van der Waals bonds between side chains of monomers | Emergence of stable conformations and semi-regular patterns in protein structures |
| Gene regulation networks | Activators and repressors | Emergence of meta-stable expression patterns |
| Cells | Membranes and channels | Emergence of compartments and cellular machinery dependent on electrochemical gradients |
| Autonomous and semi-autonomous self-replicating genetic systems | Replicator and parasite genomes | Emergence of self-nonself discrimination |
| Autonomous and semi-autonomous self-replicating genetic systems | Host cells and viruses | Emergence of infection mechanisms, defense and counter-defense systems, evolutionary arms race |
| Autonomous and semi-autonomous self-replicating genetic systems | Host cells and transposons | Emergence of intra-genomic DNA replication control; hotbeds of evolutionary innovation |
| Autonomous and semi-autonomous self-replicating genetic systems | Host cells and plasmids | Emergence of beneficial cargo genes, plasmid addiction systems, efficient gene exchange and transfer mechanisms |
| Communities of unicellular organisms | Individual cells | Emergence of information exchange and quorum sensing mechanisms; replication control apoptosis and multicellularity |
| Multicellular | Soma and germline | Emergence of complex bodies |
| organisms | | and sexual reproduction |
| Populations | Individual members | Emergence of population-level cooperation; kin selection |
| Populations | Partners with unequal parental investment (males and females) | Emergence of sexual selection and sexual dimorphism |
| Biosphere | Species in different niches | Emergence of interspecies competition, host-parasite and predator-prey relationships, mutualism |

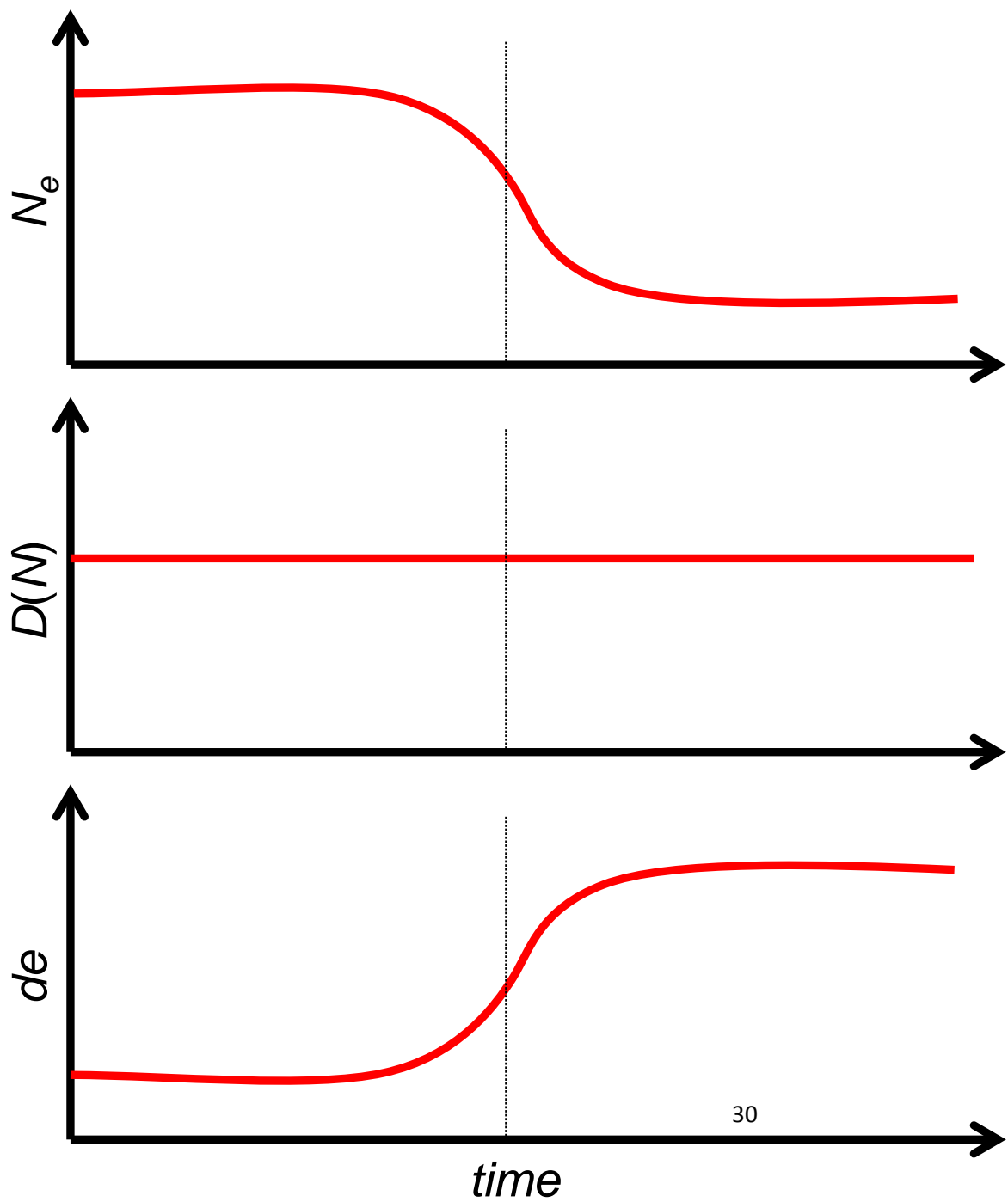

Figure 1

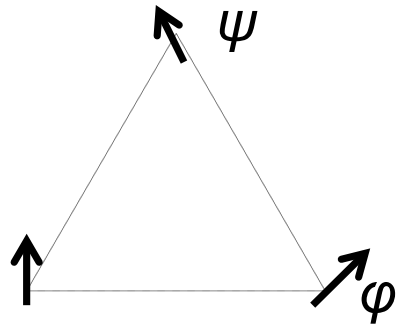
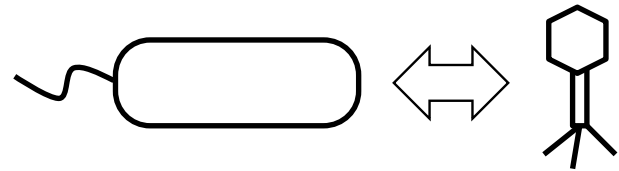
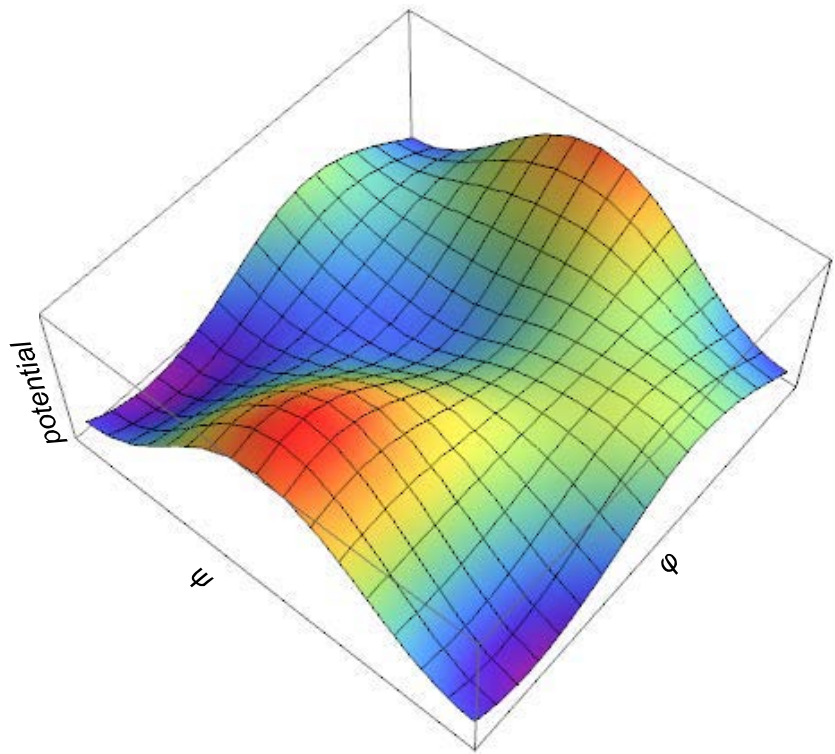
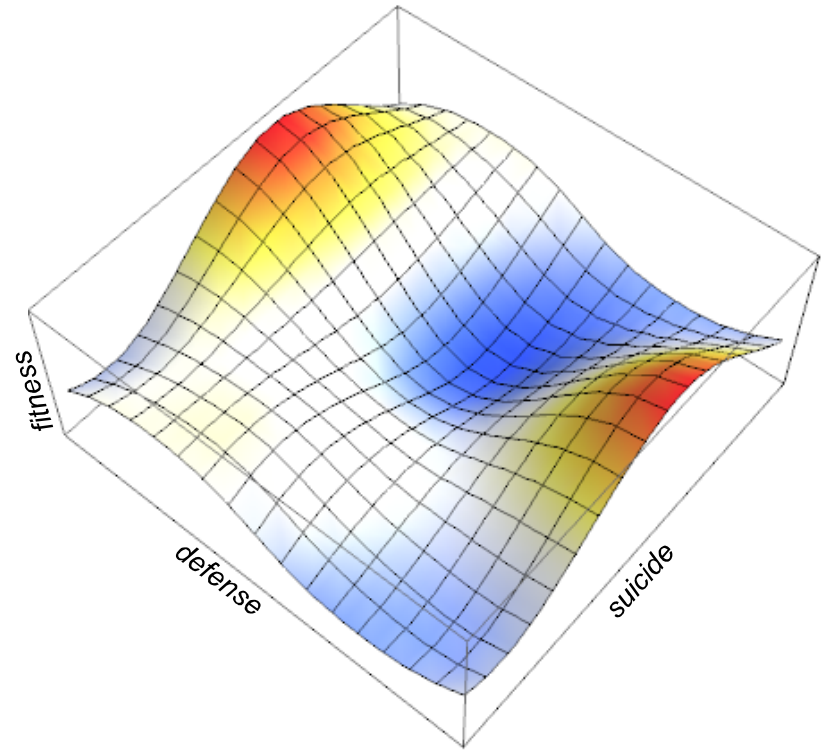

Figure 2

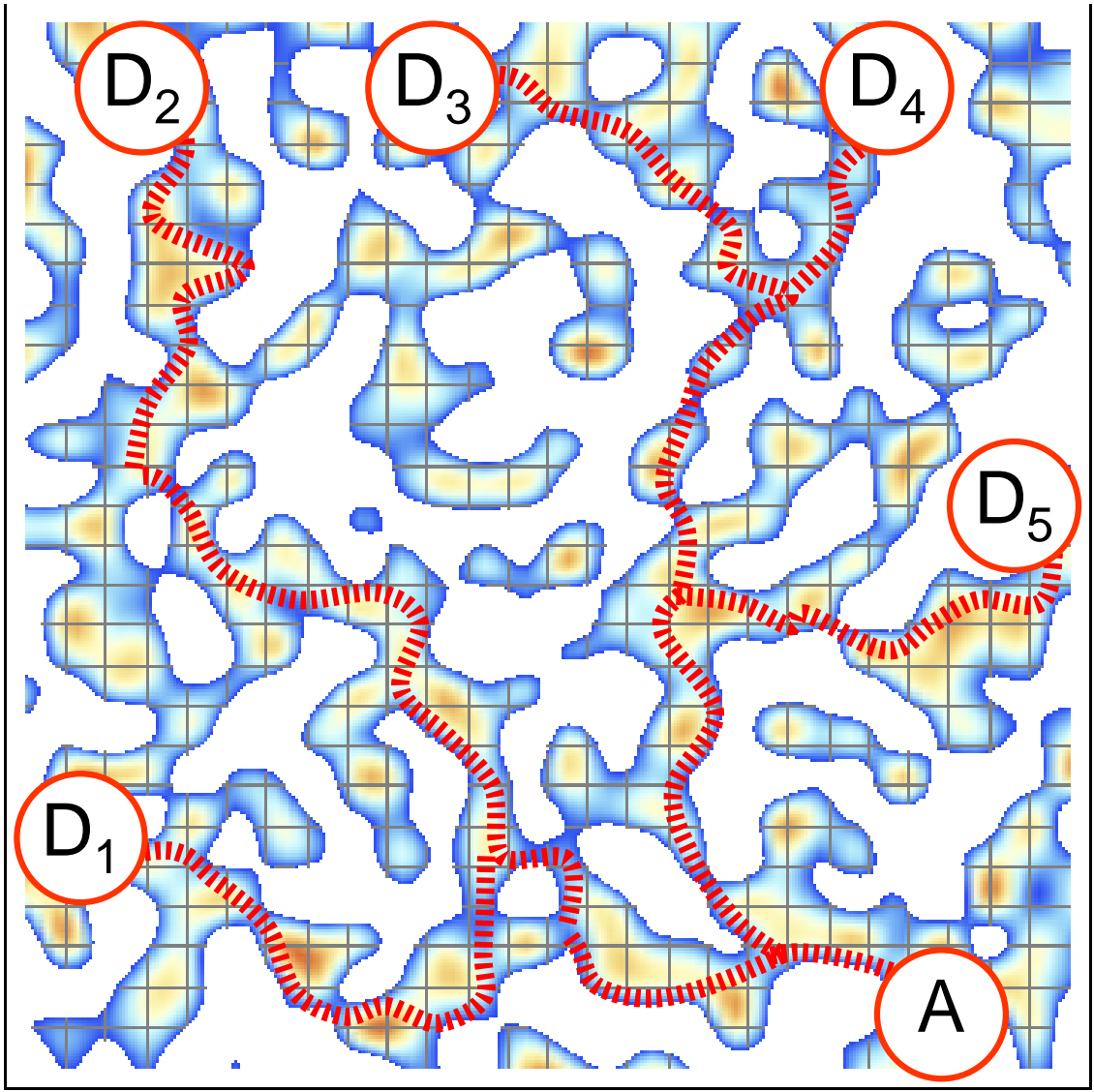

Figure 3

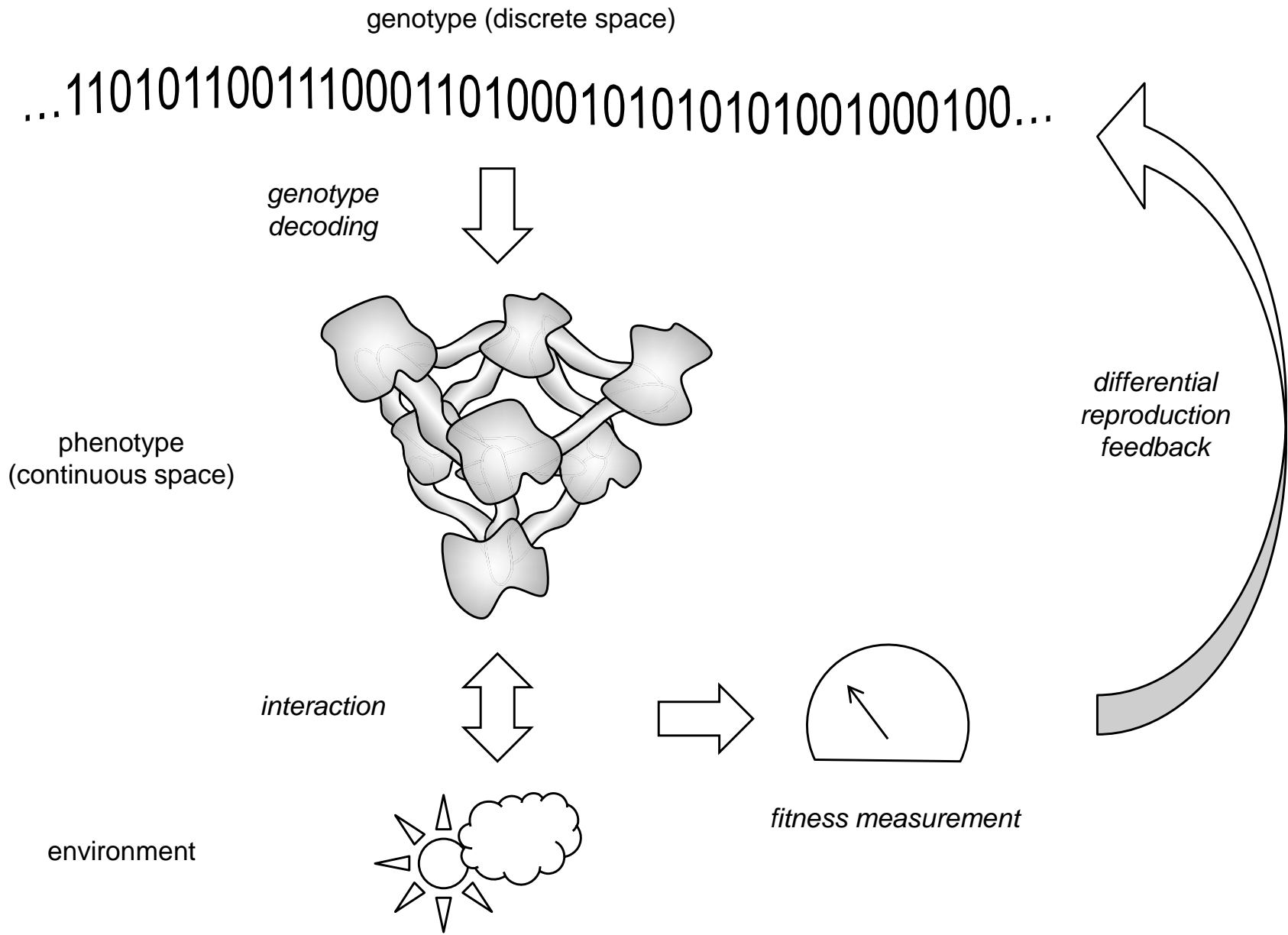

Figure 4